\documentclass[hyper]{JHEP} 

\usepackage{epsfig}

\newcommand\fverb{\setbox\pippobox=\hbox\bgroup\verb}
\newcommand\fverbdo{\egroup\medskip\noindent%
			\fbox{\unhbox\pippobox}\ }
\newcommand\fverbit{\egroup\item[\fbox{\unhbox\pippobox}]}
\newbox\pippobox
\title{Non-BPS D-brane Near  NS5-branes}

\author{by J. Kluso\v{n}\\
	 Department of Theoretical Physics and Astrophysics\\
                   Faculty of Science, Masaryk University\\
Kotl\'{a}\v{r}sk\'{a} 2, 611 37, Brno\\
Czech Republic\\
	E-mail: \email{klu@physics.muni.cz}}

\preprint{\hepth{0409298}}

\abstract{We use tachyon field theory effective action
to study  the
dynamics of a non-BPS Dp-brane propagating in the vicinity
of $k$ $NS5$-branes. For the time
dependent tachyon condensation we will concentrate
on the case of the large tachyon and the case when
a non-BPS D-brane is close to NS5-branes. 
For spatial dependent tachyon condensation we will
argue that the problem reduces to the study
of the motion of an array of D(p-1)-branes
and D(p-1)-antibranes in the vicinity
of $k$ $NS5$-branes.}


\def\bx{\mathbf{x}}
\def\mT{\mathcal{T}}
\def\bM{\mathbf{M}}

\def\bI{\mathbf{I}}
\begin{document}
\section{Introduction}\label{first}
In recent  paper 
by Kutasov \cite{Kutasov:2004dj} the problem
of the effective field theory description
of the dynamics of BPS Dp-brane in the
background of $k$ NS5-branes has
been analysed
\footnote{For other works considering
related problems,
 see \cite{Saremi:2004yd,Kutasov:2004ct,Sahakyan:2004cq,
Ghodsi:2004wn,Panigrahi:2004qr,Yavartanoo:2004wb,
Nakayama:2004yx}.}. More precisely,
Kutasov considered the stack of $k$ parallel $NS5$-branes
in type II string theory, stretched in the directions $(x^1,\dots,x^5)$
and localised in $\bx=(x^6,x^7,x^8,x^9)$
\footnote{The directions along worldvolume of
the fivebranes will be denoted by $x^{\mu} \ ,
\mu=0,1,2,\dots,5$; those transverse to
the branes will be labelled by $x^m \ , 
m=6,7,8,9$. We also use convention
$l_s=1$.}. The D-branes  that were studied
are "parallel" to the fivebranes, i.e. they are
extended in some of the fivebrane worldvolume
directions $x^{\mu}$ and pointlike in the directions
transverse to the fivebranes $(x^6,x^7,x^8,x^9)$.
Without loss of generality, we can take the worldvolume
of the Dp-brane the directions $(x^0,x^1,\dots,x^p)$.
We will label the worldvolume of the Dp-brane by $x^{\mu}$
as well, but here the index $\mu$ only runs over the range
$\mu=0,1,2,\dots,p$ with $p\leq 5$. 

Though the BPS D-branes
are stable but at the 
presence of NS5-branes
they become unstable. This follows
from the fact that configuration of parallel
$NS5$-branes and Dp-brane breaks 
supersymmetry completely. In fact, if
we place a BPS Dp-brane at finite distance
from a stack of $NS5$-branes it will experience
an attractive force and will start moving towards
to fivebranes. It was shown in \cite{Kutasov:2004dj}
that the real time dynamics of a D-brane near a stack of
$NS5$-branes exhibits a close connection to tachyon 
condensation on unstable D-brane
\cite{Sen:2002qa,Sen:2002nu,Sen:2002in,Sen:2003iv}
\footnote{For recent review of the relation 
between tachyon condensation, matrix models and 
Liouville theory, see \cite{Nakayama:2004vk} where more
extensive list of relevant papers can be found.}.
The very interesting outcome of the recent work on
real time tachyon condensation is the observation
that an effective action of Dirac-Born-Infeld (DBI)
type for the tachyon \cite{Sen:1999md,
Bergshoeff:2000dq,Garousi:2000tr,Kluson:2000iy}
captures surprisingly well many aspects of rolling
tachyon solutions of the full open string theory
\cite{Sen:2003tm,Lambert:2003zr} and is thus very useful for
studying these processes. The
origin of this agreement was partially clarified
in \cite{Kutasov:2003er,Niarchos:2004rw}, 
but see also \cite{Fotopoulos:2003yt}.
Some important questions considering
unstable D-branes in string theory were also
recently discussed in \cite{Kutasov:2004ct}.

As a next step in this research
  it seems to be natural to consider
an  unstable non-BPS Dp-brane
in the background of $NS5$-branes and
this paper is devoted to the study of this problem.  
We start in section (\ref{second})
 with the general description of parallel
unstable non-BPS Dp-brane in the background
of $k$ NS5-branes using the  form
of the tachyon effective DBI action 
proposed in 
\cite{Sen:1999md,
Bergshoeff:2000dq,Garousi:2000tr,Kluson:2000iy}.  
Then, following \cite{Kutasov:2004ct} we will
concern on the radial motion of the 
non-BPS D-brane in the background of $k$ 
coincident $NS5$-branes. We will
see that this motion depends on the
form of the tachyon condensation. 
In the first case we will 
consider the time dependent
tachyon condensation on the worldvolume
of a non-BPS Dp-brane.
We will  analyse in section (\ref{fourth}) this problem 
using   the tachyon effective action 
proposed in  
\cite{Lambert:2003zr,Kutasov:2003er,Niarchos:2004rw} 
We will generalise this action to the nontrivial
spacetime metric and dilaton. We will show 
 that in this form of the tachyon
effective action  the tachyon
field appears as  an additional embedding
coordinate and a non-BPS Dp-brane
action looks like  the DBI action of
Dp-brane embedded in eleven dimensional manifold
with specific form of the metric and dilaton
field. 
Using this form of the tachyon effective action
we will study the time dependent tachyon
condensation on the unstable D-brane in
the case when the tachyon is  large and 
non-BPS D-brane is close  to NS5-brane.
We will demonstrate that in this 
region 
of the tachyon and radial mode 
field theory space  the 
effective action posses additional global symmetry.
Then with the help of 
 the corresponding conserved charge 
we will be able to find the relation
between the tachyon and radial mode.
Then from the expression for
conserved energy 
we will get a differential equation that
describes time evolution of the radial mode
and that can be solved explicitly.

In section (\ref{fifth}) we will study the
situation when the tachyon depends on the 
 spatial coordinate on the worldvolume of
non-BPS D-brane  while the radial
 mode is function of time. In this
case the  radial mode and 
 the tachyon decouple and can be studied 
separately. In particular,  the spatial dependent
tachyon condensation 
leads to the emergence of
an array of codimension one D(p-1)-branes
and D(p-1)-antibranes 
\cite{Sen:2003tm,Kim:2003ma,Brax:2003rs,Kim:2003in}.
On the other hand the dynamics of the
radial  mode describes the collective motion of this system
and its time dependence is the same as in
the case of BPS D-brane   \cite{Kutasov:2004dj}.

Finally, in conclusion (\ref{sixth}) we outline
our results and suggest possible directions
for further research.
\section{The effective action for non-BPS
D-brane in the $NS5$-brane background}\label{second}
We restrict to the case of weak coupling, when 
$NS5$ branes are much heavier than D-branes-their
tension goes as $1/g_s^2$ while that of D-branes as
$1/g_s$. As in the paper \cite{Kutasov:2004dj} we study the
dynamics of a D-brane in vicinity of fivebranes when
we regard the fivebranes as static and study the
motion of non-BPS D-branes in their gravitation
potential. 

The background field around $k$ $NS5$-branes 
are given be the solution \cite{Callan:1991at}, where
the metric, dilaton and NS $B$-field are
\begin{eqnarray}\label{back}
ds^2=\eta_{\mu\nu}dx^{\mu}dx^{\nu}+H(x^m)
dx^mdx^m \ , \nonumber \\
e^{2(\Phi-\Phi_0)}=H(x^n) \ , \nonumber \\
H_{mnp}=-\epsilon^{q}_{mnp}\partial_q \Phi
\ . \nonumber \\
\end{eqnarray}
Here $H(x^n)$ is the harmonic function describing $k$ fivebranes
and $H_{mnp}$ is field strength of the $NS$ B-field.  For fivebranes
at generic positions $\bx_1\ , \dots \bx_k$ we have
\begin{equation}\label{harmf}
H=1+\sum_{j=1}^k\frac{1}{|\bx-\bx_j|^2} \ .
\end{equation}
For coincident fivebranes, where an $SO(4)$ symmetry
group of rotation is preserved, the harmonic function
(\ref{harmf}) reduces to 
\begin{equation}
H=1+\frac{k}{r^2} \ , r=|\bx| \ . 
\end{equation}
Our goal is to study non-BPS Dp-brane  stretched
in directions $(x^1,\dots,x^p)$. We will  label
the worldvolume of the D-brane by $\xi^{\mu} \ ,
\mu=0,\dots,p$ and use reparametrization invariance
of the worldvolume of the D-brane to set $\xi^{\mu}=x^{\mu}$.
As in case of the stable D-brane the position of 
the D-brane in the transverse directions $(x^6,\dots,x^9)$
gives to rise  to scalar fields on the worldvolume
of D-brane, $(X^6(\xi^{\mu}),\dots X^9(\xi^{\mu}))$. 
Following \cite{Sen:1999md,
Bergshoeff:2000dq,Garousi:2000tr,Kluson:2000iy}
   we can then presume
that the dynamics of the non-BPS 
Dp-brane is governed
by the action 
\begin{equation}\label{actnon}
S=-\tau_p\int  d^{p+1}\xi V(T)e^{-\Phi-\Phi_0}
\sqrt{-\det(G_{\mu\nu}+B_{\mu\nu}+
\partial_{\mu}T\partial_{\nu}T)} \ ,
\end{equation}
where $\tau_p$ is tension of non-BPS Dp-brane and 
where $V(T)$ is tachyon potential. It was conjectured
in \cite{Lambert:2003zr,Kutasov:2003er}
that this potential has the form
\begin{equation}
V(T)=\frac{1}{\cosh\frac{T}{\sqrt{2}}} \ . 
\end{equation}
We must also mention that the determinant in
(\ref{actnon}) runs over the worldvolume 
directions $\mu=0,\dots,p$, $G_{\mu\nu}$ and 
$B_{\mu\nu}$ are induced metric and $B$ field
on the non-BPS Dp-brane
\begin{eqnarray}
G_{\mu\nu}=\frac{\partial X^A}{\partial \xi^{\mu}}
\frac{\partial X^B}{\partial \xi^{\nu}}G_{AB}(X) \ ,
\nonumber \\
B_{\mu\nu}=\frac{\partial X^A}{\partial \xi^{\mu}}
\frac{\partial X^B}{\partial \xi^{\nu}}B_{AB}(X) \ , 
\nonumber \\
\end{eqnarray}
where the indices $A,B=0,\dots,9$ run over the whole ten 
dimensional spacetime so that $G_{AB}$ and
$B_{AB}$ are metric and $B$-field in ten dimensions. 

As in the paper \cite{Kutasov:2004dj} we
will consider the situation when  we place $NS5$-branes
at $\bx=0$ and restrict to the pure radial fluctuations
of non-BPS Dp-brane in the transverse $R^4$ labelled by $\bx$.
Then  the only massless field excited on the brane
is  $R(\xi^{\mu})=\sqrt{X^mX_m(\xi^{\mu})}$. This restriction
to the radial motion is consistent since, for coincident 
$NS5$-branes the background (\ref{back}) is 
$SO(4)$ invariant. Since then $B$ field is in the angular
directions and angular degrees of freedom are
not excited, the induced $B$ field vanishes. Then the
induced metric takes the form
\begin{equation}
G_{\mu\nu}=\eta_{\mu\nu}+
\partial_{\mu}X^m\partial_{\nu}X^nG_{mn}
\end{equation}
so that together with inclusion of tachyon we obtain the action
\begin{eqnarray}
S=-\tau_p\int d^{p+1}\xi
V(T)\frac{1}{\sqrt{H}}
\sqrt{-\det(\eta_{\mu\nu}+H(R)\partial_{\mu}R
\partial_{\nu}R+\partial_{\mu}T\partial_{\nu}T)} \ . 
\nonumber \\
\end{eqnarray}
We can now define the new "tachyon" field $\mT$ via the relation
\begin{equation}\label{RT}
\frac{d\mT}{dR}=\sqrt{H(R)}=
\sqrt{1+\frac{k}{R^2}} \ .
\end{equation}
In terms of this variable the non-BPS D-brane effective
action has the form
\begin{equation}\label{nonact}
S=-\int d^{p+1}\xi \mathcal{V}(T,\mT)
\sqrt{-\det(\eta_{\mu\nu}+
\partial_{\mu}\mT\partial_{\nu}\mT+
\partial_{\mu}T\partial_{\nu}T)} \ ,
\end{equation}
where we have defined
the potential $\mathcal{V}(\mT,T)$ as
\begin{equation}
\mathcal{V}(\mT,T)=\frac{\tau_p}{
\sqrt{H(R(T))}}\frac{1}{\cosh\frac{T}{\sqrt{2}}} \ . 
\end{equation}
It is interesting to analyse the asymptotic behaviour
of the potential  $\mathcal{V}$. Namely, for 
 $\mT\rightarrow -\infty$ we obtain
\begin{equation}\label{Vas}
\mathcal{V}=\tau_pe^{\frac{\mT}{\sqrt{k}}}
\frac{1}{\cosh(\frac{T}{\sqrt{2}})} \ . 
\end{equation}
 Note that for 
 $k=2$ and for  $T\rightarrow -\infty$ 
the potential (\ref{Vas}) takes the form
\begin{equation}
\mathcal{V}=2\tau_p e^{\frac{\mT}
{\sqrt{2}}+\frac{T}{\sqrt{2}}} \ .
\end{equation}
Then one can see that 
 non-BPS Dp-brane action (\ref{nonact}) 
is invariant under  $Z_2$ symmetry
\begin{equation}
Z_2: \ T'=\mT \ , \mT'=T \ . 
\end{equation}
Another enhanced $Z_2$ symmetry emerges in the limit
$ \mT\rightarrow -\infty \ , T\rightarrow \infty$ when 
(\ref{Vas}) can be written as 
\begin{equation}
\mathcal{V}=2\tau_p 
e^{\frac{\mT}{\sqrt{2}}-
\frac{T}{\sqrt{2}}} \ .
\end{equation}
It is easy to see that now the action (\ref{nonact}) is invariant
under the transformation 
\begin{equation}
T'=-\mT \ , \mT'=-T \ . 
\end{equation}
In other words for $k=2$
 one can find symmetry between $\mT$ and $T$
at asymptotic regions $\mT\rightarrow -\infty \ ,
T\rightarrow \pm \infty$.
It is also clear 
that these symmetries are broken
for finite values of $T\ , \mT$ and for $k\neq 2$.
\section{Solution of the equation of motion}
\label{third}
In this section we begin the general discussion of the solutions
of the equations of motion of the non-BPS
D-brane action (\ref{nonact}) for the case of 
coincident $NS5$-branes. 

Let us now presume that the fields
 representing the transverse position
of D-brane, $X^m \ , m=6,7,8,9$ depend only on time $X^m=
X^m(t)$.
In this case the induced $B$ field vanishes and the induced
metric takes the form
\begin{equation}
G_{\mu\nu}=\eta_{\mu\nu}+
\delta^0_{\mu}\delta^0_{\nu}
\dot{X}^m\dot{X}^mH(X^n)
\end{equation}
so that we find following action for $X^m$ and $T$
\begin{equation}\label{nonact2}
S=-\tau_p\int d^{p+1}\xi
\frac{V(T)}{\sqrt{H(X^n)}}
\sqrt{\det(\bI+\mathbf{M})} \ , 
\end{equation}
where $\bI^{\mu}_{\nu}=\delta^{\mu}_{\nu}$ is
$(p+1)\times (p+1)$ unit  matrix  and
where we have also defined  $(p+1)\times
(p+1)$   matrix
$\mathbf{M}^{\mu}_{\nu}$
as
\begin{equation}
\mathbf{M}^{\mu}_{\nu}
\equiv
\eta^{\mu\kappa}\delta_{\kappa}^0
\delta^0_{\nu}\dot{X}^m\dot{X}^m
H(X^n)+\eta^{\mu\kappa}\partial_{\kappa}T
\partial_{\nu}T \ . 
\end{equation}
Then the equations of motion for $X^m$ that arise
from (\ref{nonact2}) take the form
\begin{eqnarray}\label{Xeq}
-\frac{\partial_m H(X^n)}{2H(X^n)^{3/2}}V(T)
\sqrt{\det(\bI+\bM)}-\nonumber \\
+\frac{d}{dt}\left(\frac{1}{\sqrt{H(X^n)}}V(T)
\dot{X}^m
H(X^n)((\bI+\bM)^{-1})^{0}_{0}\sqrt{\det(\bI+\bM)}\right)+
\nonumber \\
-\frac{1}{2\sqrt{H(X^n)}}V(T)\dot{X}^n\dot{X}^n
\partial_m H(X^n)((\bI+\bM)^{-1})^{0}_{0}\sqrt{\det(\bI+\bM)}=0
\nonumber \\
\end{eqnarray}
and the equation of motion for $T$ is
\begin{eqnarray}\label{eqT}
\frac{\delta V}{\delta T}\frac{1}{\sqrt{H}}
\sqrt{\det(\bI+\bM)}-
\nonumber \\
-\partial_{\kappa}\left(\frac{V(T)}{\sqrt{H}}
\eta^{\kappa\mu}\partial_{\nu}
T((\bI+\bM)^{-1})^{\nu}_{\mu}\sqrt{\det(\bI+\bM)}
\right)=0 \ . 
\nonumber \\
\end{eqnarray}
Let us now consider the case when the
tachyon depends on time only. 
Then  the matrix $(\bI+\bM)$
is equal to
\begin{equation}
(\bI+\bM)^{\mu}_{\nu}=\left(\begin{array}{cc}
1-\dot{X}^m\dot{X}^mH(X^n)-\dot{T}^2 
& 0 \\
0 & \bI_{p\times p} \\ \end{array}\right)
\end{equation}
so that the equation of motion for $X^m$ takes the form
\begin{eqnarray}
\frac{V(T)\partial_mH(1-\dot{T}^2)}{H^{3/2}\sqrt{1
-H\dot{X}^n\dot{X}^n-\dot{T}^2}}=
\frac{d}{dt}\left(\frac{V(T)\sqrt{H}\dot{X}^m}{
\sqrt{1-H\dot{X}^n\dot{X}^n-\dot{T}^2}}\right)
\nonumber \\
\end{eqnarray}
and the equation of motion for tachyon is 
\begin{eqnarray}\label{T2}
\frac{\delta V}{\delta T}
\frac{\sqrt{1-H\dot{X}^m\dot{X}^m-\dot{T}^2}}{\sqrt{H}}
+\frac{d}{dt}
\left(\frac{V(T)\dot{T}}{\sqrt{H}\sqrt{1-H\dot{X}^m\dot{X}^m
-\dot{T}^2}}\right)=0 \ . 
\nonumber \\
\end{eqnarray}
Now we will consider the  case  
when $k$ $NS5$-branes coincide.  Then $H$ depends on 
$R=\sqrt{X^mX^m}$ and using $\partial_mH=
\frac{dH}{dR}\partial_m R=H'\frac{X^m}{R}$ we get
following equation of motion for $R$
\begin{eqnarray}\label{R}
\frac{V(T)X^mH'(1-\dot{T}^2)}{
2RH^{3/2}\sqrt{1-H\dot{X}^n\dot{X}^n-
\dot{T}^2}}=\frac{d}{dt}\left(
\frac{\dot{X}^mV(T)\sqrt{H}}{\sqrt{1-H\dot{X}^n
\dot{X}^n-\dot{T}^2}}\right)
\ , \nonumber \\
\end{eqnarray}
while the equation of motion for tachyon (\ref{T2})
does not change.  

We see that (\ref{T2}) and (\ref{R}) are complicated
differential equations of the second order. As
is well known from the study of the tachyon condensation
\cite{Sen:2002qa,Sen:2002nu,Sen:2002in,Sen:2003iv}
it is more convenient to find  all possible
 conserved charges in order to 
simplify the description of the system. The same
procedure was performed in case of BPS Dp-brane
in the NS5-brane background in \cite{Kutasov:2004dj}
where the dynamics of given Dp-brane was characterised
by conserved energy and angular momentum 
when Dp-brane moves in $(x^6,x^7)$ plane.
Now the existence of these  conserved
charges allows to analyse the possible trajectories of
Dp-brane in $NS5$-branes background
 \cite{Kutasov:2004dj}. On the other
hand for 
non-BPS Dp-brane the situation is more involved
thanks to the existence of an additional
mode $T$ on the worldvolume of a 
non-BPS Dp-brane. It seems 
to be useful   to find 
 additional conserved charge
in order to be able
study of the  dynamics of the non-BPS Dp-brane
in the fivebrane background. 
Unfortunately we were not able to find 
such a charge for general values of
$R$ and $T$. On the other hand as
we will see  in the next section 
in some region of the tachyon and radial mode
field theory space it is possible to find
this 
new conserved charge.
Then with the help of this
conservation quantity  we will be able
to determine the time evolution
$R$ and $T$ explicitly. 
\section{Another form of 
the tachyon effective action and
new conserved charge}
\label{fourth}
It turns out that in order to find additional conserved
charge it is more useful to describe the non-BPS Dp-brane
by the action  proposed in \cite{Kutasov:2003er}
 that in the 
 flat spacetime  has the form
\begin{equation}\label{Kutact}
S=\tau_p\int d^{p+1}\xi \frac{e^{-\Phi_0}}{1+\frac{T^2}{2}}
\sqrt{1+\frac{T^2}{2}+\eta^{\mu\nu}\partial_{\mu}T
\partial_{\nu}T} \ .
\end{equation}
It can be shown \cite{Kutasov:2003er} that  
there exists field redefinition that maps
(\ref{Kutact}) to (\ref{nonact}). The
existence of this redefinition  shows
that these two  actions are equivalent at least
at flat spacetime.

It is also useful   to 
rewrite (\ref{Kutact}) into the
more suggestive form 
\begin{eqnarray}\label{Kutact2}
S=-\tau_p\int d^{p+1}\xi e^{-\Phi_0} \frac{1}{
\sqrt{1+\frac{T^2}{2}}}
\sqrt{-\det(\eta_{\mu\nu}+
(1+\frac{T^2}{2})^{-1}\partial_{\mu}T\partial_{\nu}T
)}=\nonumber \\
=-\tau_p\int d^{p+1}\xi
\sqrt{F}e^{-\Phi_0}\sqrt{-\det(\eta_{\mu\nu}
+F\partial_{\mu}T\partial_{\nu}T)}
 \ ,
F=\frac{1}{1+\frac{T^2}{2}} \ . 
\nonumber \\
\end{eqnarray}
As a next step we will presume  that
 the non-BPS Dp-brane (\ref{Kutact2})
is valid in the nontrivial
closed string background as well.
Then for general spacetime metric and dilaton 
  (\ref{Kutact2}) can be written as   
\begin{equation}\label{actng}
S=-\tau_p\int d^{p+1}\xi
e^{-\Phi-\Phi_0}\sqrt{F}\sqrt{-\det(G_{\mu\nu}
+F\partial_{\mu}T\partial_{\nu}T)}
 \ .
\end{equation}
If we consider the non-BPS Dp-brane in the $NS5$-brane
background with the worldvolume parallel with the worldvolume
of fivebranes and use the static gauge 
then  the action (\ref{actng}) reduces into
\begin{equation}\label{actng2}
S=-\tau_p\int d^{p+1}\xi
\frac{\sqrt{F}}{\sqrt{H}}\sqrt{-\det(\eta_{\mu\nu}
+H\partial_{\mu}X^m\partial_{\nu}X^m
+F\partial_{\mu}T\partial_{\nu}T)}
 \ .
\end{equation}
If we now forget about  the origin of the tachyon 
then  we can consider
the action  (\ref{actng2}) 
as  the DBI action for Dp-brane embedded
in  nontrivial eleven dimensional spacetime 
where the additional
coordinate is labelled by $T$ and where
the metric and dilaton 
are given as
\begin{eqnarray}
ds^2=dx_{\mu}dx^{\mu}+H(x^{\mu})
dx^mdx^m+F(T)dT^2 \ , 
\nonumber \\
e^{2(\Phi-\Phi_0)}=\frac{H(x^n)}{F(T)} \ .
\nonumber \\
\end{eqnarray}
It would be certainly very nice to have such a 
geometrical interpretation of the tachyon on the worldvolume
of unstable D-brane, especially in the
context of the recent paper \cite{Kutasov:2004ct}.
Of course more work is needed to prove
whether  this conjecture is true. 

Let us now 
consider   the radial motion of D-brane in
the background of $k$ coincident $NS$-branes.
By $SO(4)$ rotation-symmetry of the problem, we
can consider the moving of non-BPS Dp-brane
in two dimensional plane, say the $(x^6,x^7)$.
Then it is useful to introduce polar coordinates
\begin{equation}
X^6=R\cos\theta \ ,
X^7=R\sin \theta \ . 
\end{equation}
At the moment let us now presume that
$\theta$ is zero which, as
we will see, corresponds to the zero angular momentum.
Then  the  action (\ref{actng2})
takes the form
\begin{equation}\label{actng2a}
S=-V_p\tau_p\int dt \frac{1}{
\sqrt{1+\frac{T^2}{2}}
\sqrt{1+\frac{k}{R^2}}}
\sqrt{1-(1+\frac{k}{R^2})\dot{R}^2
-(1+\frac{T^2}{2})^{-1}\dot{T}^2} \ . 
\end{equation}
When $T^2/2\gg 1$ and $k/R^2\gg 1$
the action (\ref{actng2a}) simplifies considerably 
\begin{equation}\label{actng3}
S=-V_p\tau_p\int dt \frac{1}{
\sqrt{\frac{T^2}{2}\frac{k}{R^2}}}
\sqrt{1-\frac{k}{R^2}\dot{R}^2
-\frac{2}{T^2}\dot{T}^2} \  .
\end{equation}
It is easy to see that (\ref{actng3}) 
 is invariant under the transformation
\begin{equation}
T'=\lambda T \ , R'=\lambda R \ ,
\end{equation}
where  $\lambda$ is constant.  As usual this symmetry
implies an existence of the conserved charge. In fact, for
$\lambda=1+\epsilon \ ,\epsilon\ll 1$ we get
\begin{equation}\label{varTR}
\delta T=\epsilon T \ , \delta R=\epsilon R
\end{equation}
Since the action is invariant for constant $\epsilon$ then
for  $\epsilon=\epsilon(t)$ its variation should be
proportional to the time derivation of $\epsilon$
\footnote{For more detailed discussion
of conserved charges and Noether method
in the context of string theory, see
\cite{Sen:2004yv,Sen:2004cq}.}
\begin{equation}
\delta S=\int dt J\dot{\epsilon}=
-\int dt\dot{J}\epsilon \ .
\end{equation}
For fields obeying equation of motion $\delta S=0$
and we obtain the conserved charge $\dot{J}=0$.
In the case of the variations (\ref{varTR}) the
corresponding conserved charge is
\begin{eqnarray}\label{J}
J=V_p\tau_p\sqrt{\frac{2}{k}}\frac{R}{T}\left(
\frac{k\dot{R}}{R}+\frac{2\dot{T}}{T}\right)
\frac{1}{\sqrt{1-\frac{k}{R^2}\dot{R}^2
-\frac{2}{T^2}\dot{T}^2}} \ . \nonumber \\
\end{eqnarray}
The second conserved charge that will be useful
for the study of the dynamics of the non-BPS Dp-brane
is the energy  
\begin{equation}\label{E}
E=P\dot{R}+\Pi \dot{T}-L=
V_p\tau_p\sqrt{\frac{2}{k}}
\frac{R}{T}\frac{1}{\sqrt{1-\frac{k}{R^2}\dot{R}^2
-\frac{2}{T^2}\dot{T}^2}} \ , 
\end{equation}
where 
\begin{eqnarray}
P=\frac{\delta L}{\delta \dot{R}}=
V_pk\tau_p\sqrt{\frac{2}{k}}
\frac{\dot{R}}{TR\sqrt{1-\frac{k}{R^2}\dot{R}^2
-\frac{2}{T^2}\dot{T}^2}} \ , \nonumber \\
\Pi=\frac{\delta L}{\delta \dot{T}}=
2V_p\tau_p\sqrt{\frac{2}{k}}\frac{R\dot{T}}
{T^3\sqrt{1-\frac{k}{R^2}\dot{R}^2
-\frac{2}{T^2}\dot{T}^2}} \ . 
\nonumber \\
\end{eqnarray}
Since $J$ and $E$ 
contain an overall volume factor $V_p$ it is more
natural to work with the densities (which we
label with the same letters $E,J$) when we 
strip of the volume factors. 
From (\ref{J}) and (\ref{E})  we can easily
find the relation between $R$ and $T$
\begin{eqnarray}\label{TR}
\frac{J}{E}=
\frac{k\dot{R}}{R}+\frac{2\dot{T}}{T}=
k\frac{d\ln R}{dt}+2\frac{d\ln T}{dt}=
\frac{d}{dt}\left(\ln R^k+\ln T^2\right)
\nonumber \\
\frac{J}{E}=\frac{d\ln R^kT^2}{dt} 
\Rightarrow 
 R^kT^2=C^2e^{\frac{J}{E}t} 
\Rightarrow
T=Ce^{\frac{J}{2E}t}R^{-k/2} \ ,  
\nonumber \\
\end{eqnarray}
where the integration constant
$C$ will be determined below using
the value of the tachyon $T_0$ at
time $t_0=0$. 
Using  (\ref{TR}) we also get  
\begin{equation}\label{dotTR}
\dot{T}=(\frac{J}{2E}-\frac{k\dot{R}}{2R})T \ . 
\end{equation}
Now if we insert (\ref{dotTR}) together with (\ref{TR})
into (\ref{E}) we obtain differential equation
with parameters $E$ and $J$  
that determines the time evolution of $R$.
 Let us now consider
the situation  when $J=0$. Then (\ref{TR}) implies
\begin{equation}\label{TRJ0}
\dot{T}=
-\frac{k}{2}\frac{\dot{R}}{R}T \ . 
\end{equation}
Inserting (\ref{TRJ0}) into (\ref{E}) we obtain 
the differential equation for $R$
\begin{eqnarray}
\dot{R}^2=\frac{1}{(k+\frac{k^2}{2})}
\left(R^2-\frac{2\tau_p^2}{kE^2C^2}R^{4+k}
\right) \ 
\end{eqnarray}
that has the solution 
\begin{equation}\label{ueq}
\frac{t}{\sqrt{k+\frac{k^2}{2}}}
=\mp\frac{2}{2+k}\mathrm{arctanh}\sqrt{
1-\frac{2\tau_p^2}{kE^2C^2
}R^{2+k}}+C_0 \ . 
\end{equation}
If we demand that at 
$t=0$ the non-BPS Dp-brane
is in  its turning point 
we obtain that $C_0$ is equal to zero
and consequently 
\begin{eqnarray}\label{Rt}
 \frac{1}{R}=\left(\frac{2\tau_p^2}{kE^2C^2}\right)^{\frac{1}{2+k}}
\left(\cosh\left(
\frac{(2+k)t}{2\sqrt{k+k^2/2}}\right)\right)^{\frac{2}{2+k}}
 \ .
\nonumber \\
\end{eqnarray}
We see that the time dependence of the radial coordinate
is different from the  case of BPS Dp-brane studied 
in \cite{Kutasov:2004dj}.
In fact, for $t\rightarrow \infty$  we get
\begin{equation}\label{scalR}
R\sim e^{-\frac{t}{\sqrt{k+\frac{k^2}{2}}}} \ (
\mathrm{non-BPS \ D-brane}) \ ,
R\sim e^{-\frac{t}{\sqrt{k}}} \ 
(\mathrm{BPS \ D-brane}) \ . 
\end{equation} 
Physical   explanation of  
this difference is as follows. 
The effective tension of non-BPS 
D-brane lowers in   the process of the tachyon condensation.
Consequently the gravitative
attraction from the NS5-brane becomes weaker than
in the case of BPS D-brane and hence non-BPS
D-brane approaches NS-brane slowly than
BPS D-brane. 

In order to determine the time dependence of $T$ we
use (\ref{TRJ0}) that together with (\ref{Rt})
gives   
\begin{equation}\label{Tt}
T=CR^{-k/2}=C\left(\cosh\left(
\frac{(2+k)t}{2\sqrt{k+k^2/2}}\right)\right)^{\frac{k}{2+k}}
\left(\frac{2\tau_p^2}{kE^2C^2}\right)
^{\frac{k}{2(2+k)}} \ .
\end{equation}
If we demand that 
 at $t=0$ the tachyon is equal to 
 $T_0\gg 1$ then using  
(\ref{Tt}) we can find relation between
$T_0$,  energy $E$  and the integration constant $C$ 
\begin{equation}
C=T_0^{\frac{2+k}{2}}
\left(\frac{kE^2}{2\tau_p^2}\right)^{\frac{k}{4}} \ .
\end{equation}
It is also interesting to study the
time dependence of the components of
 the stress energy tensor. To begin
with let us replace the 
flat worldvolume 
metric $\eta_{\mu\nu}$ 
with the general one $g_{\mu\nu}$  in the action
(\ref{actng2}). Then we get
\begin{equation}\label{actng3t}
S=-\tau_p\int d^{p+1}\xi\sqrt{-g}
\frac{\sqrt{G}}{\sqrt{H}}\sqrt{\det(\delta^{\mu}_{\nu}
+Hg^{\mu\kappa}\partial_{\kappa}X^m\partial_{\nu}X^m
+Fg^{\mu\kappa}
\partial_{\kappa}T\partial_{\nu}T)}
 \ .
\end{equation}
Since the action has the form $S=-\int d^{p+1}
\xi \sqrt{-g}\mathcal{L}$ then
 the stress energy tensor defined as
$T_{\mu\nu}=-\frac{2}{\sqrt{-g}}\frac{\delta S}
{\delta g^{\mu\nu}}$ is equal to
\begin{equation}
T_{\mu\nu}=-g_{\mu\nu}\mathcal{L}+
2\frac{\delta \mathcal{L}}{\delta g^{\mu\nu}} \ .
\end{equation}
After insertion of the flat spacetime metric and
using the fact that the worldvolume fields are
the time dependent only we get 
\begin{eqnarray}
T_{00}=\frac{\tau_p\sqrt{F}}{\sqrt{H}}
\frac{1}{\sqrt{1-H\dot{X}^m\dot{X}^m-F\dot{T}^2}} \ , \nonumber \\
T_{ij}=-\delta_{ij}\frac{\tau_p\sqrt{F}}{\sqrt{H}}
\sqrt{1-H\dot{X}^m\dot{X}^m-F\dot{T}^2} \  \nonumber \\
\end{eqnarray}
 that for large $T$ and small $R$ take
the forms
\begin{eqnarray}
T_{00}=\tau_p\sqrt{\frac{2}{k}}\frac{R}{T}
\frac{1}{\sqrt{1-\frac{k\dot{R}^2}{R^2}-
\frac{2\dot{T}^2}{T^2}}} \ , \nonumber \\
T_{ij}=-\delta_{ij}\tau_p\sqrt{\frac{2}{k}}
\frac{R}{T}
\sqrt{1-\frac{k\dot{R}^2}{R^2}-
\frac{2\dot{T}^2}{T^2}} \ . \nonumber \\
\end{eqnarray}
Firstly we know   that $T_{00}=E$ is constant.
On the other hand the
 spatial components $T_{ij}$ are equal to
\begin{equation}
T_{ij}=-\delta_{ij}\frac{\tau_p2R}{T}
\sqrt{1-(k+\frac{k^2}{2})\frac{\dot{R}^2}{R^2}}\sim
\frac{e^{-\frac{k+2}
{2\sqrt{k+k^2/2}}t}}{\sqrt{t}}\rightarrow 0 \ .
\end{equation}
We see that 
as in the case of the BPS D-brane \cite{Kutasov:2004dj} the pressure
goes exponentially to zero at asymptotic future.
This asymptotic behaviour  is
also in agreement with the rolling tachyon description
\cite{Sen:2002qa,Sen:2002nu,Sen:2002in,Sen:2003iv}. 

Now we will briefly discuss the case  of nonzero $J$. We
can think about this as follows. Inserting  
(\ref{dotTR}) into (\ref{E})
we obtain the equation 
\begin{equation}\label{ss}
(k+\frac{k^2}{2})\dot{R}^2=
R^2(1-\frac{J^2}{2E^2})+\frac{Jk\dot{R}R}{2E}
-\frac{2\tau_p^2}{kE^2C^2}R^{4+k} \ . 
\end{equation}
It is very difficult to find general solution of
 this equation so that we restrict ourselves
to the case of large $t$. Then by presumption 
$R\ll 1$ and we can  neglect the term $R^{4+k}$
in (\ref{ss}).  If we 
now presume the time dependence of $R$ for $t\rightarrow
\infty $ is 
$R\sim e^{\beta t}$ then we get from (\ref{ss}) the
quadratic equation for $\beta$ 
\begin{eqnarray}\label{betas}
(k+\frac{k^2}{2})\beta^2-\frac{Jk}{2E}\beta-(1-\frac{J^2}{2E^2})=0
\Rightarrow  \nonumber \\
\beta=\frac{1}{(k+\frac{k^2}{2})} \left(-\sqrt{k+\frac{k^2}{2}}
\sqrt{(1-\frac{J^2}{2E^2})+\frac{J^2k^2}{4E^2
(k+k^2/2)}}+\frac{Jk}{2E} \right)\ . \nonumber \\
\end{eqnarray}
We see that  (\ref{betas}) is  equal to
$-\frac{1}{\sqrt{k+\frac{k^2}{2}}}$ for
$J=0$ with agreement with the scaling
(\ref{scalR}).  More precisely,      let us presume 
that $\frac{J}{E}\ll 1$. Then $\beta$ given in
(\ref{betas}) can be written as
\begin{equation}
\beta\approx-\frac{1}{\sqrt{k+\frac{k^2}{2}}}
+\frac{Jk}{E(k+\frac{k^2}{2})}
 \ . 
\end{equation}
As in the case of the vanishing charge $J$, \
 $T_{00}$ is energy density equal to $E$ 
while the spatial components of the stress
energy tensor have asymptotic form
\begin{eqnarray}
T_{ij}\approx-\delta_{ij}e^{-\left(\frac{2+k}
{2\sqrt{k+\frac{k^2}{2}}}-\frac{(2+k)J}{2E(k+\frac{k^2}{2})}\right)t} \ . 
\nonumber \\
\end{eqnarray}
We again see that 
the pressure vanishes at far future. This result
confirms the presumption that the time evolution of
the system at nonzero $J$ has similar form as
in the case of $J=0$. 

Finally  we will consider the case of general
$\theta$. Now the 
action (\ref{actng3}) is equal to 
\begin{equation}\label{actng3L}
S=-V_p\tau_p\int dt \frac{1}{
\sqrt{\frac{T^2}{2}\frac{k}{R^2}}}
\sqrt{1-\frac{k}{R^2}\left(
\dot{R}^2+\dot{\theta}^2R^2\right)
-\frac{2}{T^2}\dot{T}^2}
\end{equation}
It is again easy to see 
that this action is invariant under
scaling $T'=\lambda T \ ,   
R'=\lambda R$. The  corresponding
conserved charge is equal to 
\begin{equation}
J=V_p\tau_p\sqrt{\frac{2}{k}}\frac{R}{T}
\left(\frac{k\dot{R}}{R}+\frac{2\dot{T}}{T}\right)
\frac{1}{\sqrt{1-\frac{k}{R^2}(\dot{R}^2+R^2\dot{\theta}^2)
-\frac{2}{T^2}\dot{T}^2}} \ .
\end{equation}
We have also the conserved energy
\begin{equation}\label{EL}
E=V_p\tau_p\sqrt{\frac{2}{k}}
\frac{R}{T}\frac{1}{\sqrt{1-\frac{k}{R^2}(
\dot{R}^2+R^2\dot{\theta}^2)-\frac{2}{T^2}
\dot{T}^2}} \ . 
\end{equation}
One can also see that the action (\ref{actng3L}) is invariant
under the transformation $\theta'=\theta+\epsilon$, where
$\epsilon$ is constant. The corresponding
conserved charge is angular momentum and
it is equal to
\begin{eqnarray}\label{Lc}
L=V_p\tau_p\sqrt{\frac{2}{k}}
\frac{R}{T}\frac{k\dot{\theta}}
{\sqrt{1-\frac{k}{R^2}(
\dot{R}^2+R^2\dot{\theta}^2)-\frac{2}{T^2}
\dot{T}^2}} \ .  \nonumber \\
\end{eqnarray}
Now we can solve this system of equations for fixed
$E \ , L \ , J$
\footnote{Now $E \ , J \ , L$ mean 
corresponding densities.}. For simplicity we restrict ourselves
to the case $J=0$ which allows us to express $T$ as function
of $R$ exactly in the same way as in the
case of $J=0$ given
in (\ref{TRJ0}). Inserting this expression into (\ref{Lc})
we can express $\dot{\theta}$ as function of
$R \ , \dot{R}$ 
\begin{eqnarray}
\dot{\theta}^2=\frac{L^2\left(1-(k+\frac{k^2}{2})
\frac{\dot{R}^2}{R^2}\right)}{
\left(L^2k+\frac{2\tau_p^2kR^{2+k}}{C^2}
\right)} \ . 
\nonumber \\
\end{eqnarray}
If we insert this  result into  (\ref{EL}) and using
(\ref{TRJ0}) we obtain the differential 
equation for $R$ 
\begin{equation}
(k+\frac{k^2}{2})\dot{R}^2=
\frac{(kE^2-C^2L^2)}{kE^2}
R^2-\frac{2\tau_p^2}{kE^2C^2}R^{4+k} 
\end{equation}
with the general solution 
\begin{equation}
\frac{t}{\sqrt{k}E\sqrt{k+k^2/2}}
=\mp\frac{2}{(2+k)\sqrt{kE^2-C^2L^2}}
\mathrm{arctanh}\sqrt{1-\frac{2\tau_p^2}
{kE^2C^2-C^4L^2}R^{2+k}}+C_0 \ . 
\end{equation}
If we demand that at 
 $t=0$ non-BPS Dp-brane is in
its turning point we get $C_0=0$. 
 Then we finally obtain
\begin{eqnarray}\label{Rtx}
\frac{1}{R^{2+k}}=\frac{2\tau_p^2}
{kE^2C^2(1-\frac{C^2L^2}{kE^2})}\cosh\left(
\frac{2+k}{2\sqrt{k+\frac{k^2}{2}}}\sqrt{1-\frac{C^2L^2}{kE^2}}t
\right) \ . 
\nonumber \\
\end{eqnarray}
We see that nonzero angular momentum slows down the exponential
decrease of $R$ as $t\rightarrow \infty$. This is the same behaviour
as was observed in case of 
 BPS Dp-brane in \cite{Kutasov:2004dj}. Since 
\begin{equation}
\frac{L}{E}=k\dot{\theta}
\end{equation}
one can easily find time dependence of $\theta$
\begin{equation}\label{thetat}
\theta=\frac{L}{Ek}t \ . 
\end{equation} 
The solution (\ref{Rtx}),(\ref{thetat}) and
the relation $T=CR^{-k/2}$ describes the non-BPS
D-brane that during the worldvolume tachyon condensation
spirals towards to the origin, circling around it an infinite
number of times in the process. 

In this section we have studied the time dependent
tachyon condensation on non-BPS Dp-brane in
the $NS5$-brane background in the regime where
$k/R^2\gg 1$ and $T^2\gg 1$. 
In the next section we will consider the second
example of the tachyon condensation
when the tachyon is spatial dependent while the
radial coordinate depends on time only.
\section{Spatial dependent tachyon}\label{fifth}
In this section we will analyse the case when the
tachyon depends on one spatial coordinate,
say $\xi^{1}=x$. We will study this problem
using the non-BPS Dp-brane action given
in (\ref{nonact2}). For this configuration
the matrix  $\bI+\bM$ takes the form
\begin{equation}
(\bI+\bM)^{\mu}_{\nu}=
\left(\begin{array}{ccc}
1-\dot{X}^m\dot{X}^mH & 0 & 0  \\
0 &1+ T'^2  & 0 \\
0 & 0 & \bI_{(p-1)\times (p-1)} \\ 
\end{array}\right) \ , 
\end{equation}
where  $T'\equiv \frac{dT}{dx}$. 
Consequently the
action (\ref{nonact2}) simplifies as
\begin{eqnarray}\label{LTs}
S=-\int dt \mathcal{L}=-V_{p-1}\int  
dt dx\frac{V(T)}{\sqrt{H}}
\sqrt{\det(\bI+\bM)}=\nonumber \\
=-V_{p-1}\int dt dx \frac{V(T)}{\sqrt{H}}
\sqrt{1-H\dot{X}^m\dot{X}^m}\sqrt{
1+T'^2} \ , \nonumber \\
\end{eqnarray}
where $V_{p-1}$ is volume of $p-1$ dimensional
spatial section of the worldvolume of Dp-brane. 
The equation of motion for $X^m$  that follow
from (\ref{LTs}) are
\begin{eqnarray}\label{eqXs}
V(T)\sqrt{1+T'^2}\left[-\frac{\partial_m H}
{2H^{3/2}\sqrt{1-H\dot{X}^n
\dot{X}^n}}+
\frac{d}{dt}\left(\frac{\sqrt{H}\dot{X}^m}
{\sqrt{1-H\dot{X}^n\dot{X}^n}}\right)\right]=0
\nonumber \\
\end{eqnarray}
using the fact that $T$ does not depend on $t$. 
The equation of motion for the tachyon is
\begin{equation}\label{eqTs}
\frac{\sqrt{1-H\dot{X}^n\dot{X}^n}}
{\sqrt{H}}\left[
\frac{\delta V}{\delta T}\sqrt{1+T'^2}-\frac{d}{dx}
\left(\frac{V(T)T'}{\sqrt{1+T'^2}}\right)\right]=0
 \ ,
\end{equation}
where now we have used 
the fact that $X^m$ are functions of $t$ only. 
From (\ref{eqXs}) and (\ref{eqTs}) we see that 
 tachyon and scalar modes decouple
 and can be studied separately. 

The solution considering the spatial dependent tachyon is
well known 
\cite{Sen:2003tm,Lambert:2003zr,
Kim:2003ma,Brax:2003rs,Kim:2003in}.
In particular, for
$V=\frac{\tau_p}{\cosh(\frac{T}{\sqrt{2}})}$ 
the solution of  (\ref{eqTs}) takes the form 
\begin{eqnarray}
\sinh\frac{T}{\sqrt{2}}=
\sqrt{\frac{\tau_p^2}{K^2}-1}\sin \frac{x}{\sqrt{2}} \ ,
\nonumber \\
\end{eqnarray}
where $K$ is integration constant. 
The corresponding energy density is
\begin{equation}
\rho(x)=V(T)\sqrt{1+T'^2}=\frac{V^2(T)}{K}=
\frac{\tau_p^2}{K}\frac{1}
{1+\left(\frac{\tau_p^2}{K^2}-1\right)
\sin^2\frac{x}{\sqrt{2}}} \ .
\end{equation}
The interpretation of this solution is usually in
terms of array of kinks and antikinks. Then by
integrating the energy density 
over a half period of the solution we can find energy
of the kink
\begin{equation}
T_{p-1}\equiv \int_{-\sqrt{2}\pi/2}
^{\sqrt{2}\pi/2}dx \rho(x)=\pi\sqrt{2}\tau_p
\end{equation}
This is nothing but the tension of the BPS kink
identified as D(p-1)-brane. Therefore the tachyon solution
may be interpreted as representing an array of 
D(p-1)-branes and D(p-1)-antibranes in the background of
$k$ $NS5$-branes. The fact that we consider 
$X^m$'s independent on $x$ implies that 
$X^m$'s parametrise the collective motion of the
configuration of D(p-1)-branes and D(p-1)-antibranes. 

Now let as return to the study of the dynamics of
$X^m$, where we will specialise to the case of 
coincident
fivebranes. As in the previous subsection we will
demand that $H$ only depends on $R=\sqrt{X^nX^n}$ so that
the equation of motion (\ref{eqXs}) takes the form
\begin{equation}
\frac{d}{dt}\left(\frac{\dot{X}^m\sqrt{H}}
{\sqrt{1-H\dot{X}^n\dot{X}^n}}\right)=
\frac{H'X^m}{
2RH^{3/2}\sqrt{1-H\dot{X}^n
\dot{X}^n}} \ , 
\end{equation}
where now $H'\equiv \frac{dH}{dR}$. In order to
solve this equation we must specify the initial data
$X^m(t=0)$ and $\dot{X}^m(t=0)$. As was argued
in \cite{Kutasov:2004dj}
 these two vectors define plane
in the transverse $R^4$ that can be by $SO(4)$ rotation
taken to the $(x^6,x^7)$ plane. 
Then the motion
will remain in this plane for all time.
 More precisely, the angular momentum
for the motion in $(x^6,x^7)$ plane
can be determined
 through the
Noether method
 with 
the result
\begin{equation}
L=V_{p-1}\int dx V(T)\sqrt{1+T'^2}
\frac{\sqrt{H}(X^6\dot{X}^7-\dot{X}^6X^7)}
{\sqrt{1-H\dot{X}^n\dot{X}^n}} \ . 
\end{equation}
In the same way we can determine the conserved
energy from the requirement of the invariance of 
the action with respect to the time translation
\begin{equation}
E=V_{p-1}\int dx \left(P_m\dot{X}^m
-\mathcal{L}\right)=V_{p-1}\int dx\frac{V(T)\sqrt{1+T'^2}}{\sqrt{H}
\sqrt{1-H\dot{X}^n\dot{X}^n}}
\end{equation}
using
\begin{equation}
P_m(x)=\frac{\delta \mathcal{L}}{\delta \dot{X}^m}=
V_{p-1}\frac{V(T)\sqrt{H}\dot{X}^m\sqrt{1+T'^2(x)}}
{\sqrt{1-H\dot{X}^n\dot{X}^n}}
\end{equation}
We again introduce the coordinate 
\begin{equation}
X^6=R\cos\theta \ ,
X^7=R\sin \theta 
\end{equation}
so that the conserved energy and momentum 
are equal to 
\begin{eqnarray}\label{en}
E=V_{p-1}\int dx \frac{V(T)\sqrt{1+T'^2}}
{\sqrt{H}\sqrt{1-H(\dot{R}^2-R^2\dot{\theta}^2)}} \ ,
\nonumber \\
L=V_{p-1}\int dx \frac{V(T)\sqrt{1+T'^2}R^2\dot{\theta}}
{\sqrt{H}\sqrt{1-H(\dot{R}^2+R^2\dot{\theta}^2)}} \ . 
\nonumber \\
\end{eqnarray}
We have argued above that the tachyon condensation has
physical interpretation as an array of D(p-1)-branes and 
D(p-1)-antibranes. In the same way we know that
  $R,\theta$ do not depend on
$x$ so that it is natural to write 
\begin{equation}
E=(N+\overline{N})E_{p-1} \ , 
L=(N+\overline{N})L_{p-1} \ ,
\end{equation}
where
\begin{equation}
(N+\overline{N})=\frac{\int_{-\infty}^{\infty} dx \rho(x)}
{\int_{-\sqrt{2}\pi/2}^{\sqrt{2}\pi/2}dx\rho(x)} \ 
\end{equation} 
is number of D(p-1)-branes and D(p-1)-antibranes. 
Then it is natural to  introduce an energy  and
angular momentum densities per single
D(p-1)-brane or D(p-1)-antibrane through
the relations
\begin{equation}
e\equiv\frac{E\int_{-\sqrt{2}\pi/2}^{\sqrt{2}\pi/2}dx\rho(x)}
{V_{p-1}\int dx \rho(x)} \ ,
l\equiv\frac{L\int_{-\sqrt{2}\pi/2}^{\sqrt{2}\pi/2}dx\rho(x)}
{V_{p-1}\int dx \rho(x)} \
\end{equation}
that using (\ref{en})
are equal to
\begin{eqnarray}\label{el}
e= \frac{T_{p-1}}
{\sqrt{H}\sqrt{1-H(\dot{R}^2+R^2\dot{\theta}^2)}} \ ,
\nonumber \\
l=\frac{T_{p-1}R^2\dot{\theta}^2}
{\sqrt{H}\sqrt{1-H(\dot{R}^2+R^2\dot{\theta})}} \ ,
\nonumber \\
\end{eqnarray}
where now $T_{p-1}$ is tension of BPS D(p-1)-brane
or D(p-1)-antibrane. 
Using the second equation in (\ref{el}) we find
\begin{equation}\label{thetadot}
\dot{\theta}^2=\frac{l^2(1-H\dot{R}^2)}
{\frac{T^2_{p-1}}{H}R^4+l^2R^2}
\end{equation}
and hence we get
\begin{equation}
\dot{R}^2=\frac{1}{H}-\frac{1}{e^2H^2}
\left(T_{p-1}^2+\frac{l^2}{R^2}\right)
\end{equation}
Let us consider the case of vanishing momentum
$l=0$. Then (\ref{thetadot}) implies that $\theta$ is
constant   and the equation for radial coordinate
becomes
\begin{equation}\label{dotR2}
\dot{R}^2=\frac{1}{H}-
\frac{T_{p-1}^2}{e^2H^2} \ .
\end{equation}
If now the trajectory remains in the small $R$ region, such that
$R\ll \sqrt{k}$ for all times which happens for $T_{p-1}/e\ll 1$
we can solve for the trajectory $R(t)$ exactly since in this
case $H(R)=k/R^2$ and we get
\begin{equation}
\dot{R}^2=\frac{R^2}{k}-\frac{T_{p-1}^2R^2}
{e^2k^2}
\end{equation}
with the solution
\begin{equation}
\frac{1}{R}=\frac{T_{p-1}}{e\sqrt{k}}
\cosh \frac{t}{\sqrt{k}} \ . 
\end{equation}
In summary, the spatial dependent tachyon condensation
on the non-BPS Dp-brane leads to the array of
D(p-1)-branes and D(p-1)-antibranes. If we do not worry
about stability  of this configuration 
then  this problem reduces to the
description of the motion of  D(p-1)-brane
or D(p-1)-antibrane in the background of
$k$ $NS5$-branes.   
\section{Conclusion}\label{sixth}
This paper was devoted to the study of
the dynamics of the non-BPS Dp-brane in the
background of $k$ NS5-branes, where
we have proceed 
in the similar way as in the case of
BPS D-brane that was studied in \cite{Kutasov:2004dj}. In 
other words, we have analysed 
the dynamics of the non-BPS Dp-brane
using the effective field theory 
description
that is based on the DBI-like tachyon
effective action. We have considered
two possible cases of the tachyon 
condensation.  In the first case,
which was time dependent tachyon
condensation, 
we have shown that 
in the approximation of large tachyon potential
and in the region  very close to the
$k$ $NS5$ coincident branes
a non-BPS Dp-brane 
approaches 
the worldvolume of fivebranes with exponentially
growing tachyon and exponentially
decreasing radial mode where the
factor in the exponential function
 was different from
the factor given in the case of BPS Dp-brane.
We have explained this difference as a
result of the tachyon condensation on
the worldvolume of non-BPS Dp-brane
that effectively reduces the tension of
D-brane. We have also shown that
the pressure exponentially vanishes
with the agreement with the paper  \cite{Kutasov:2004dj}.
 We would
like to stress that these results were
obtained thanks 
 to the existence 
of  an additional conserved charge whose
existence follows from the enhanced 
symmetry of the tachyon effective action
for large $T$ and small $R$. 

At this place however we must stress one
important point considering the
validity of the effective field
theory description of the
non-BPS D-brane. 
The DBI-like tachyon 
effective action was not directly obtained 
from the first principles of string theory,
even if its success in the description
of some aspects of the tachyon 
condensation 
is intriguing. On the other hand it was
recently stressed in 
\cite{Fotopoulos:2003yt} 
that it is not completely clear how to 
interpret the action (\ref{actnon}). 
It is also well known 
that there exist other form of the tachyon
effective actions 
\cite{Minahan:2000tf,Gerasimov:2000zp,
Kutasov:2000aq,
Kutasov:2000qp,Tseytlin:2000mt} 
that are generally valid in 
different regions 
of the tachyon field theory space. It is 
commonly believed that these different 
tachyon effective actions are related by
complicated field redefinition. It would
be  very interesting to see whether these
different tachyon effective actions can
be studied in the nontrivial background
as well. 
Another problem that deserves to be studied
is the analysis of the dynamics of the
non-BPS D-brane in the higher dimensional
D-brane background. 
The situation is now
more involved thanks to the nontrivial
Ramond-Ramond fields that 
 couple
to the worldvolume of non-BPS D-brane
\cite{Okuyama:2003wm,Billo:1999tv}.
We hope to return to this problem in
future. 
\\
\\
{\bf Acknowledgement}

This work was supported by the
Czech Ministry of Education under Contract No.
14310006.
\\

\end{document}